\documentclass[useAMS,usenatbib,onecolumn]{mn2e}
\usepackage{graphicx}
\usepackage[utf8x]{inputenc}
\title[Composition and evolution of Grain mantle]
{Composition and evolution of Interstellar Grain mantle under the effects of photo-dissociation}
\author[Ankan Das, Sandip K. Chakrabarti]
{Ankan Das$^{1}$, Sandip K. Chakrabarti$^{2}$\\
$^{1}$Indian Centre for Space Physics, Chalantika 43, Garia Station Road,
             Kolkata 700084, India\\
$^{2}$S. N. Bose National Centre for Basic Sciences, Salt Lake,
              Kolkata 700098, India}
\begin{document}

\date{}

\maketitle

%\label{firstpage}

\begin{abstract}
We study the chemical evolution of interstellar grain mantle by varying the physical parameters of the 
interstellar medium (ISM). To mimic the exact interstellar condition, gas grain 
interactions via accretion from the gas phase and desorption (thermal evaporation and 
photo-evaporation) from the grain surface are considered. We find that the 
chemical composition of the interstellar grain mantle is 
highly dependent on the physical parameters associated with a molecular cloud. 
Interstellar photons are seen to play an important role towards the growth and the structure of the interstellar
grain mantle. We consider the effects of interstellar photons (photo-dissociation and 
photo-evaporation) in our simulation under various interstellar conditions.
We notice that the effects of interstellar photons dominate around the region of lower visual extinction.
These photons contribute significantly in the formation of the grain mantle. Energy of the incoming photon is
attenuated by the absorption and scattering by the interstellar dust. Top most layers are 
mainly assumed to be affected by the incoming radiation. We have studied the effects of photo-dissociation
by varying the number of layers which could be affected by it. Model calculations are 
carried out for the static (extinction parameter is changing with the density of the cloud) 
as well as the time dependent case (i.e., extinction parameter and number density of the cloud both are 
changing with time) and the results are discussed in details. Different routes to the formation of water 
molecules are studied and it is noticed that around the dense region, production of water molecules 
via O$_3$ and H$_2$O$_2$ contributes significantly. At the end, various observational evidences for 
the condensed phase species are summarized with their physical conditions and are compared with our 
simulation results.

\end{abstract}


\section{Introduction}
The densities in the interstellar space (consisting of 99\% gas and 1\% dust), 
even in the densest interstellar clouds, are 
so low that they are beyond the capabilities of the best vacuum systems on Earth. 
Even then, a large number of infrared absorption features are observed toward low
and high mass protostars. Some of these features are attributed to the
solid state interstellar molecules \citep{Boo04}. Discovery of these interstellar molecules in 
condensed phase demands a proper explanation for the interstellar chemical composition budget.
Increasing observational and experimental evidences 
in the gas phase points to the inadequacy of the gas phase chemistry and
suggests the need to incorporate the grain surface processes. 
Interstellar dust grains are thought to be consisting of an amorphous
silicate or carbonaceous core surrounded by a molecular ice layer \citep{Draine03, Gibb04}.
Presently, very little is known about the morphology or chemistry on these grain surfaces, 
or the porosity of the grains themselves.
Similarly, the chemical and physical processes on the grain surfaces till date are also very poorly known.
A number of theoretical attempts have been carried out to verify the observed 
abundances of the gas phase and the condensed phase species \citep{Holl70,Wat72,Allen75,
Tiel82,Hase92,Charn01,Stan02,Cup07,Chak06a,Chak06b,Das08a,Das08b,Das10}. Both the
deterministic and stochastic approaches are tested by these studies. Observational and experimental evidences suggest 
that almost 90\% of the grain mantle is covered with H$_2$O, CH$_3$OH and CO$_2$. 
In our earlier paper \citet{Das10}, (hereafter DAC10), 
we mentioned that it is very difficult to theoretically match the exact observed abundances 
of all the three molecules simultaneously. Results of the numerical simulations show that only in a 
narrow region of parameter space all these three molecules are produced within the observed limit. 

It is well known that the ISM chemistry is highly influenced by the interstellar UV radiations which
mainly come from the nearby young stars. UV radiation affects the chemistry of the ISM
through photo-dissociation and photo-ionization of the molecules and atoms. 
Several works are carried out to study the photo-chemistry in the ice containing simple as well as complex molecules. 
Most of the studies \citep{Hagen79, Dhen82, Alla88} are concentrated following the irradiation of ice mixtures 
to reflect the chemical composition of the star forming regions. These types of studies give us an
insight about the photo effects on the interstellar ice mantles.

Structure of the interstellar ices under various physical circumstances are yet to 
fully understood. Water ice is less abundant along some lines of sight. In the present paper, we 
carry out a coupled gas-grain study to fully understand the structure of the interstellar 
grain mantle around various lines of sight. Gas-grain interactions are considered via 
accretion from gas phase and evaporation from grain surface. Important surface reactions, which are
feasible under the interstellar conditions are considered in our surface chemistry network to
have clear insight of the interstellar ice feature. Moreover, to mimic the realistic situation,
the formation of several mono-layers are considered.
As we are interested only around the dense region of the molecular cloud, we assume that the major
portion of the gas phase carbon atoms are already converted into carbon monoxide and as 
in the other models \citep{Stan02,Das08b,Das10}, we also consider the accretion of 
H, O and CO from the gas phase to follow the time evolution of several chemical species.
Recent experimental findings of several new reaction pathways and several interaction energies involved during these process,
motivate us to update our earlier DAC10 model. Our main aim in the present context is to
study the formation of ice mantle under the influence of photo-processing and to
make a comparative study among different routes to the water formation. This has been suggested
recently by some experiments \citep{Ioppolo08}. We also compare our simulation results with 
other observational, experimental and theoretical results carried out till date.
In Section 2, we discuss the physical processes and the method used for the modeling.
Extensive discussions along with the results are presented in Section 3.  
In Section 4, a comparison between the theoretical and observational results are presented.
Finally, in Section 5 we have summarized our work.

\section{Procedure}
In general, the rate equation method \citep{Hase92,Robert02} is widely used 
to follow the surface process occurring under the
interstellar condition. Main advantage of this method is that the computation is very fast. 
This method assumes the average concentration of species to follow the chemical evolution.
Main drawback with this method is that it does not
consider the statistical fluctuations and discrete nature of the species. As a result, this
method is very much dependent on the physical conditions of the interstellar cloud. 
In the lower accretion rate regime, where surface population is comparatively lower, this method 
tends to overestimate the result \citep{Chak06a,Chak06b,Das08b, Das10}. Moreover as this method deals with the average 
concentration, it is not able to follow the internal structure of the grain mantle. However, as the time
passes by, grains are expected to grow in size with several surface species. As the rate 
equation is unable to trace any particular species, it is not possible for this method
to mimic the actual situation in ISM. This is the reason, we use the descrete-time random walk Monte 
Carlo method to have more accurate results. In our approach, fluctuations in the surface abundance 
due to the statistical nature of the 
grain is preserved and as a result this method is much more realistic to study the interstellar
surface chemistry. With this method, we can follow movements of each species 
and evaluate the results at every instant.

Details of the processes we follow are already discussed in \citet{Das10}. For the sake of completeness, 
we discuss them again briefly. 
We assume that the grains are square shaped and  similar to an atom on an fcc[100] plane, each grain site has 
four nearest neighbors. We assume the periodic boundary conditions 
and also allow the possibility to form multiple layers on a grain surface.
For modeling purposes, we consider classical dust grains, characterized by 1000A$^{\circ}$ and 10$^6$ surface sites.
Following \citet{Hase92}, we consider the surface density of available sites to be 1.5 $\times 10^{15}$ 
cm$^{-2}$. Number density of the grains are taken to be 1.33$ \times 10^{-12} n$ 
($n$ is the concentration of H nuclei in all forms) and gas to dust ratio is taken to be $100$ by mass\citep{Hase92}. 
Monte Carlo method is computationally much slower than the rate equation method and it is 
quite difficult to simulate for a grain having $10^6$ sites. 
To save computational time, we consider the grain having $50 \times 50$ sites
and the final results are extrapolated for the grain having $10^6$ sites. Reason behind choosing the grain 
having $50\times50$ sites is that it is above the limit of the statistical fluctuating 
size and according to \citet{Chang05}, above this size, results are scalable and do not differ significantly.
Location of the accreting species is dictated by generating a pair of random numbers.
Langmuir Hinshelwood (reactions between the surface species via hopping) and Eley Rideal (reaction between the
incoming species with the surface species), both types of reactions are considered. After landing on a grain, 
its direction towards any one of the four nearest neighbors is dictated by generating  random numbers.
Two possibilities may occur: If the site is occupied then either it
will react when the reaction is permitted or it will wait until its next hopping time. 
If the neighboring site is vacant and it does not have any species just beneath that site then it can roll down 
on that grid until it touch some existing species in a layer or just 
the bare grain surface. If the reaction between the species touching the existing species is 
permitted, a new species is formed and if not, then it can sit on the top of that. 
For the reactions having barriers, we are generating  random numbers upon each encounter and
checking for the possibility of the reaction. Following \citet{Hase92} probability (P) for the reaction to happen upon
an encounter is calculated and if the generated random number is less than P  then the reaction is allowed and if 
the generated number is greater than P then reaction is discarded at that instant.
Every species has its own hopping time scale. In our simulation, we consider the hopping 
time scale of H atom to be the minimum. After each interval of
H hopping time, we are checking for the evaporation (via thermal or other process) and
photo-dissociation by generating random numbers. For example, let us consider H$_2$O. Its evaporation
rate is R$_{ev}$ sec$^{-1}$. That means, one H$_2$O can evaporate after each $\frac{1}{R_{ev}}$ 
seconds of its formation on the grain. Since the smallest time scale is the hopping time scale $t_H$
of an H atom, after every $t_H$, we generate a random number to find out 
whether the species will evaporate or stick to the grain. If the generated number is smaller 
than $t_H \times R_{ev}$, then the species can evaporate. Photo-dissociations are also treated 
similarly, i.e., if the generated random number is under 
the probability of photo-dissociation of any species, that species 
can photo-dissociate. Gas phase compositions are adjusted at every $t_H$ after taking care of
the accretion onto  and the evaporation from the grains.

A major difference between our previous study and the 
present study is that we have updated our reaction network by considering several new reactions
and have considered the effects of the cosmic rays as well. Cosmic rays have the direct/induced 
effect on the interstellar grain mantle. In our model, we consider the photo-dissociation 
and photo evaporation to incorporate the effects of the cosmic rays.
Formation of the most abundant molecules such as, H$_2$O, CH$_3$OH and CO$_2$ are
considered as before. Keeping in mind the recent experimental results \citep{Ioppolo08}, we 
have also included the formation of H$_2$O via O$_3$ and H$_2$O$_2$ routes.
All the reactions considered in our present model are given in Table 1. 
Activation barrier energy used in our previous paper \citep{Das10}, 
and in some other papers \citep{Stan02,Cup07} are noted in Tab.1.
We gave some comments on the most recent experimental/theoretical findings in the last column.
For the reaction numbers 4 and 6, we normally use $2000$K as the activation barrier 
energy \citep{Stan02}. However, the most recent findings by \citet{Fuchs09}, 
suggest that the energy barrier could be much lower ($390 \pm 40$K for reaction number 4
and $415\pm 40$K for the reaction number 6 at $12$K). In our model calculation, we assume that
at $10$K, the energy barrier is $390$K for the reaction number 4, and $415$K for the reaction number 6. 
So far, barrier energy $1200$K was used as mentioned in \citet{Mel79}.
Recent experimental findings by \citet{Ioppolo08} suggest that the reaction number 12 is barrier less.
Recent surface chemical modelling by  \citet{Cazau10} assumed that the reaction number 14 is also barrier less 
though it is not experimentally verified.
In our model, we assume reaction number 12 is barrier less and reaction number 14 with barrier 1400K \citep{Klem75}.
Reaction number 16 is considered to have the activation energy $2600$K \citep{Schiff73}.
Binding energies of the species are the keys for the chemical enrichment of the interstellar 
grain mantles. Grain surface provides the space for the interstellar gas phase species to land on
and to react with the other surface species. If allowed, chemical species can return back to 
the gas phase after chemical reactions or in the original accreted form. 
Reactions between the surface species is highly dependent on the mobility of the surface
species, which in turn, depends on the thermal hopping time scale or on the
tunneling time scale. For the lighter species, tunneling time scale is much shorter 
than the hopping time scale. So for the lighter species, surface migration by tunneling method
might be effective. When two surface species meet, they can react. If some activation energy involved
for that reaction to happen, they can react with a quantum mechanical tunneling probability (P). 
\citet{Goumans10} studied the tunnelling reaction between O and CO with
direct dynamics employing density functional theory and they obtained the activation energy
2500K for the reaction between O and CO. We use only thermal hopping of O atom for its mobility on the grain surface
and use the same activation barrier energy, 2500K following \citet{Goumans10} for the reaction number 9 of Tab. 1.
In \citet{Das08b}, a comparative study between tunneling and hopping for the mobility of H atom on the grain surface 
was carried out and it was noticed that results did not show drastic variation within the 
accretion regime considered here (number density 10$^3$-10$^6$). 
We expect that as long as the accretion rate is within the limit , this does not  
significantly affect the final results.

There are two types of binding energies associated with the 
physical processes at low temperature, namely, physisorption and chemisorption. 
Among them, physisorption energies are more probable in low temperature ($\sim 10$K) condition, 
i.e., frigid (low temperature, $\sim$ 10K) condition. Therefore, in our model calculations, we considered the physisorption
energies. We defined the whole system with a lot of interaction energies. 
Most of the energies considered here are discussed in detail in \citet{Das10} and references therein.
The rest of the species which are updated in our present model are given in Table 2. 
It is very difficult to find the exact binding energies for every specific surface.
In \citet{Das10}, olivine surface was considered for the modeling purpose. 
Presently, we also considered olivine surface as a model substrate.
Among the updated chemical species, binding energies of HO$_2$, O$_3$ and H$_2$O$_2$ are taken from 
\citet{Cazau10}. They used the physisorbed energies of these molecules for the 
Carbonaceous grain and here we have included their values as if those are the interaction energies
with the bare olivin surfaces. For C atom, we used the same binding energy as in CO. For 
CH$_3$ and CH$_2$OH, we used the same binding energies as in CH$_3$OH and H$_2$CO respectively. 
Since the binding energies of these species are very high, we can be sure that 
our assumptions would not alter the final results significantly.
Due to the lack of knowledge of the appropriate interaction energies between various species, 
we assumed that the binding energies of these species with other surface species  
are the same as with olivine surfaces.
\begin{table}
\centering
\caption{Surface Reactions considered in our network}
\begin{tabular}{|l|l|l|l|}
\hline
& Reactions & E$_a$(K)             & Comments\\
\hline
1  & H+H $\rightarrow$ H$_2$          &    &  \\
2  & H+O $\rightarrow$ OH             &    &  \\
3  & H+OH $\rightarrow$ H$_2$O        &    &  \\
4  & H+CO $\rightarrow$ HCO           & 2000& 390K $\pm$ 40K at 12K \citep{Fuchs09} \\
5  & H+HCO $\rightarrow$ H$_2$CO      &     & \\
6  & H+H$_2$CO $\rightarrow$ H$_3$CO  & 2000& 415K $\pm$ 40K at 12K \citep{Fuchs09} \\
7  & H+H$_3$CO $\rightarrow$ CH$_3$OH &     & \\
8  & O+O $\rightarrow$ O$_2$          &     & \\
9 & O+CO $\rightarrow$ CO$_2$        & 1000 & 2970K \citep{Talbi06}, 2500K \citep{Goumans10}\\
10 & O+HCO $\rightarrow$ CO$_2$+H     &     & \\
11 & O+O$_2\rightarrow$ O$_3$         &     &  \\ 
12 & H+O$_2\rightarrow$HO$_2$         & 1200& barrier less \citep{Ioppolo08}\\
13 & H+HO$_2\rightarrow$H$_2$O$_2$         &&  \\
14 & H+H$_2$O$_2\rightarrow$H$_2$O+OH       &  1400& barrier less \citep{Cazau10}\\
15 & H+O$_3\rightarrow$ O$_2$+OH         &450&  \\
16 & H2+OH$\rightarrow$ H$_2$O+H&2600         &  \\
\hline
\end{tabular}
\end{table}
\begin{table*}
\begin{center}
\addtolength{\tabcolsep}{-3pt}
%\centering
{\tiny
\caption{Energy barriers in degree Kelvin}
\begin{tabular}{|c|c|c|c|c|c|c|c|c|c|c|c|c|c|c|c|c|c|c|c|c|}
\hline
{\bf Species} & \multicolumn{19}{|c|}{\bf Substrate}\\
\hline
%&\multicolumn{2}{|c}}{silicate}&H&$H_2$&O&$O_2$&OH&$H_2O$&CO&HCO&$H_2CO$&$H_3CO$&$CH_3OH$&$CO_2$\\

&\bf{Silicate}&\bf{H}&$\bf{\mathrm{H_2}}$&\bf{O}&$\bf{\mathrm{O_2}}$&\bf{OH}&$\bf{\mathrm{H_2O}}$&\bf{CO}&\bf{HCO}&$\bf{\mathrm{H_2CO}}$&$\bf{\mathrm{H_3CO}}$&$\bf{\mathrm{CH_3OH}}$&$\bf{\mathrm{CO_2}}$&$\bf{\mathrm{HO_2}}$&$\bf{\mathrm{H_2O_2}}$&\bf{$\mathrm{O_3}$}&\bf{$\mathrm{C}$}&\bf{$\mathrm{CH_3}$}&\bf{$\mathrm{CH_2OH}$}\\
\hline
\bf{H}&350&350&45&350&45&350&650&350&350&350&350&350&350&350&350&350&350&350&350\\
\hline
\bf{$\mathrm{H_2}$}&450&30&23&30&30&30&440&450&450&450&450&450&450&450&450&450&450&450&450\\
\hline
\bf{O}&800&480&55&480&55&55&800&480&480&480&480&480&480&800&800&800&800&800&800\\
\hline
\bf{$\mathrm{O_2}$}&1210&69&69&69&69&1000&1210&1210&1210&1210&1210&1210&1210&1210&1210&1210&1210&1210&1210\\
\hline
\bf{OH}&1260&1260&240&240&240&240&3500&1260&1260&1260&1260&1260&1260&1260&1260&1260&1260&1260&1260\\
\hline
\bf{$\mathrm{H_2O}$}&1860&390&390&390&390&390&5640&1860&1860&1860&1860&1860&1860&1860&1860&1860&1860&1860&1860\\
\hline
\bf{CO}&1210&1210&1210&1210&1210&1210&1210&1210&1210&1210&1210&1210&1210&1210&1210&1210&1210&1210&1210\\
\hline
\bf{HCO}&1510&1510&1510&1510&1510&1510&1510&1510&1510&1510&1510&1510&1510&1510&1510&1510&1510&1510&1510\\
\hline
\bf{$\mathrm{H_2CO}$}&1760&1760&1760&1760&1760&1760&1760&1760&1760&1760&1760&1760&1760&1760&1760&1760&1760&1760&1760\\
\hline
\bf{$\mathrm{H_3CO}$}&2170&2170&2170&2170&2170&2170&2170&2170&2170&2170&2170&2170&2170&2170&2170&2170&2170&2170&2170\\
\hline
\bf{$\mathrm{CH_3OH}$}&2060&2060&2060&2060&2060&2060&2060&2060&2060&2060&2060&2060&2060&2060&2060&2060&2060&2060&2060\\
\hline
\bf{$\mathrm{CO_2}$}&2500&2500&2500&2500&2500&2500&2500&2500&2500&2500&2500&2500&2500&2500&2500&2500&2500&2500&2500\\
\hline
\bf{$\mathrm{HO_2}$}&2160&2160&2160&2160&2160&2160&2160&2160&2160&2160&2160&2160&2160&2160&2160&2160&2160&2160&2160\\
\hline
\bf{$\mathrm{H_2O_2}$}&2240&2240&2240&2240&2240&2240&2240&2240&2240&2240&2240&2240&2240&2240&2240&2240&2240&2240&2240\\
\hline
\bf{$\mathrm{O_3}$}&2900&2900&2900&2900&2900&2900&2900&2900&2900&2900&2900&2900&2900&2900&2900&2900&2900&2900&2900\\
\hline
\bf{$\mathrm{C}$}&1210&1210&1210&1210&1210&1210&1210&1210&1210&1210&1210&1210&1210&1210&1210&1210&1210&1210&1210\\
\hline
\bf{$\mathrm{CH_3}$}&2060&2060&2060&2060&2060&2060&2060&2060&2060&2060&2060&2060&2060&2060&2060&2060&2060&2060&2060\\
\hline
\bf{$\mathrm{CH_2OH}$}&1760&1760&1760&1760&1760&1760&1760&1760&1760&1760&1760&1760&1760&1760&1760&1760&1760&1760&1760\\
\hline
\end{tabular}}
\vskip 0.2cm
\scriptsize{For the all species $E_b=0.3\ E_D$ is used except the atomic hydrogen where
$E_b=0.2857 \ E_D$ is used. For the reference please see the text.}
\label{table-2}
\end{center}
\end{table*}
\begin{table*}
\centering
\caption{Interstellar photo reactions considered at the present work.}
\begin{tabular}{|l|l|l|l|l|l|l|l|}
\hline
& &\multicolumn{3}{|c|}{Cosmic ray induced rates (R$_i$)}&\multicolumn{3}{|c|}{Direct cosmic ray rates (R$_d$)} \\
\cline{3-8}
& Reactions &$\alpha$ &$\beta$&$\gamma$&$\alpha$&$\beta$&$\gamma$ \\
\hline
1 &$ ^a$O$_2\rightarrow$ O+O         &1.0(-16)&0(0)&3.755(2)&1.90(-9)&0(0)&1.85(0)  \\
2 & $^a$OH$\rightarrow$ O+H         &1.0(-16)&0(0)&2.545(2)&3.90(-10)&0(0)&2.24(0) \\
3 & H$_2$O$\rightarrow$ H+OH         &1.3(-17)&0(0)&4.855(2)&5.90(-10)&0(0)&1.70(0) \\
4 & CO$\rightarrow$ C+O         &1.3(-17)&0(0)&1.050(2)&2.00(-10)&0(0)&2.50(0) \\
5 & HCO$\rightarrow$ H+CO         &1.3(-17)&0(0)&2.105(2)&1.10(-9)&0(0)&8.0(-1) \\
%6 & H$_2$CO$\rightarrow$ H+HCO         &&&&&& \\
6 & H$_2$CO$\rightarrow$ H$_2$+CO         &1.3(-17)&0(0)&1.329(3)&7.00(-10)&0(0)&1.7(0) \\
7 & CH$_3$OH$\rightarrow$ CH$_3$+OH         &1.3(-17)&0(0)&7.520(2)&6.00(-10)&0(0)&1.8(0) \\
%9 & CH$_3$OH$\rightarrow$ CH$_2$OH+H         &&&&&& \\
%10 & CH$_3$OH$\rightarrow$ H$_3$CO+H         &&&&&& \\
8 & CO$_2$$\rightarrow$ CO+O         &1.3(-17)&0(0)&8.540(2)&1.40(-9)&0(0)&2.50(0) \\
9 & H$_2$O$_2$$\rightarrow$ OH+OH         &1.3(-17)&0(0)&7.500(2)&5.90(-10)&0(0)&1.7(0) \\
10 & $^a$O$_3$$\rightarrow$ O$_2$+O         &1.0(-16)&0(0)&3.755(2)&1.95(-9)&0(0)&1.85(0) \\
11 & $^a$HO$_2$$\rightarrow$ OH+O         &1.0(-16)&0(0)&3.750(2)&6.70(-9)&0(0)&2.12(0) \\
\hline
\end{tabular}
\vskip 0.2cm
\scriptsize{All rates are taken from \citet{Wood07} except the reactions denoted by $^a$ \citep{Cazau10}.}
\end{table*}

\section{Results \& Discussion}
\subsection{Effect of photo dissociation}
Photo-dissociation of the interstellar grain mantle plays an important role in the 
evolution of the interstellar grain mantle. This feature is included in our present model.
Due to the lack of knowledge about the different photo-dissociation rates of the surface species,
we assumed that the photo-dissociation rates of the surface species are the same 
as in the gas phase species. List of photo reactions along with the various parameters necessary for the 
direct cosmic ray photo reactions and cosmic ray induced photo reactions 
are noted in Table 3. Reactions which are taken from \citet{Wood07}, rates are calculated by;
\begin{equation}
R_i =\alpha (\frac{T}{300})^{\beta} \frac{\gamma}{1-\omega} \ \ s^{-1} ,
\end{equation}
where, $\alpha \ , \ \beta \ , \ \gamma$ are three constants, 
$T$ is the grain temperature, $\omega$ is the dust grain albedo 
in far ultra-violet. Here we use $\omega=0.6$ in our all calculations. 
For the other Cosmic ray induced photo reactions which are marked by
`$^a$' in Table 3, we calculate the rate by the following equation;
\begin{equation}
R_i=\alpha \gamma .
\end{equation}
For the direct interstellar photo reactions, the rate is derived as;
\begin{equation}
R_d=\alpha exp(-\gamma A_v) \ \ s^{-1} ,
\end{equation}
where, $A_v$ is the extinction at visible wavelengths caused by the interstellar dust.

Normally, first few layers are assumed to be affected by the interstellar photons. Keeping this in mind,
we use Monte-Carlo approach to investigate this aspect. Probabilities of 
photo reactions and photo evaporations are applied by the use of random number generators. 
After the photo dissociation, the heavier counterpart is allowed to replace the dissociated species
and the comparatively lighter counterpart is stored in the nearby vacant sites or 
react with other nearby suitable species of the same layer. If there are no site available on the
same layer to satisfy the criterion, dissociated lighter species may drop down to the lower vacant site
or act with some other suitable reactant. After this, if
it fails to find out a suitable location then the lighter species  is allowed to 
evaporate. We notice that during our simulation time scale, the evaporation probability of the 
lighter counterpart is close to zero, so this assumption should not result in any error. 
Photo-evaporation are also incorporated in our code by the help of the random number generator as discussed earlier. 

Mobility of a species inside the ice leads to chemical enrichment of the
interstellar grain mantle. Photo-dissociation adds up a new dimension to the
ice mantle by changing its morphology. 
%Though the first few layers are mainly 
%affected by the effect of the direct Cosmic ray photo-dissociation or cosmic ray induced photo-dissociation. 
Its contribution towards the final structure of the interstellar grain mantle should not be neglected 
around some regions of the ISM.

\begin {figure}
\vskip 2.0cm
\centering{
\includegraphics[width=7cm]{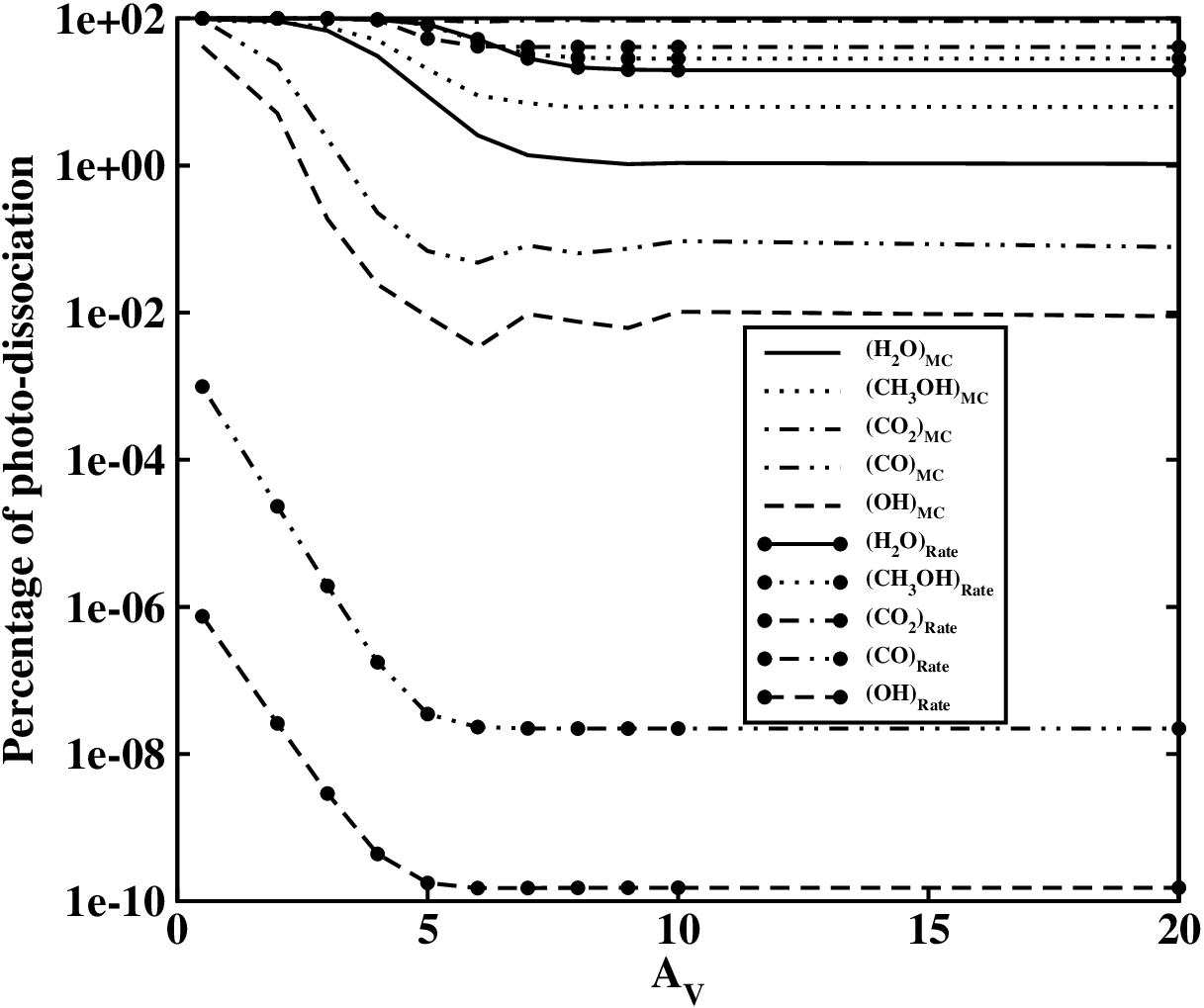}}
\caption{Variation of percentage of photo-dissociation  as a function of the extinction parameter A$_V$.
Results obtained by Monte-Carlo simulations are found to be consistently lower compared to those obtained from
rate equations.}
\label{fig-2}
\end {figure}

\subsubsection{Dependence on the Extinction parameter}

In Fig. 1, percentage of photo-dissociated molecules are plotted with the
extinction parameter. Both the Rate equation and Monte Carlo methods are considered for a thorough study of these effects.
Along the `Y' axis we have shown the percentage of molecules  which are photo-dissociated  for a given species.
For the better agreement of our simulated results with the observations, we choose the initial conditions around the 
favorable zone mentioned in DAC10 \citep{Das10}. DAC10 parameters consist of hydrogen number density 
($\sim 10^4$ cm$^{-3}$), atomic Hydrogen number density ($\sim 1.10$ cm$^{-3}$), atomic Oxygen number density 
($\sim 1.05$ cm$^{-3}$), Carbon-monoxide number density ($0.375$ cm$^{-3}$), and temperature 
($10$K). Unless otherwise stated, throughout the paper we use the DAC10 parameters as the initial condition 
and all the results are shown after evolving the system up to two million years. For the regions having 
other number densities, we scale the number density of O and CO accordingly by keeping the atomic 
hydrogen number density of $1.10$ cm$^{-3}$. It is evident from Fig. 1 that  the
lower extinction regions are heavily affected by the interstellar photons. For this simulation, we assume that only 
first mono-layer of the evolving grain mantle is exposed to the interstellar gas and
photo-reaction/evaporation can only occur from the first mono-layer.
\begin {figure}
\vskip 2.0cm
\centering{
\includegraphics[width=7cm]{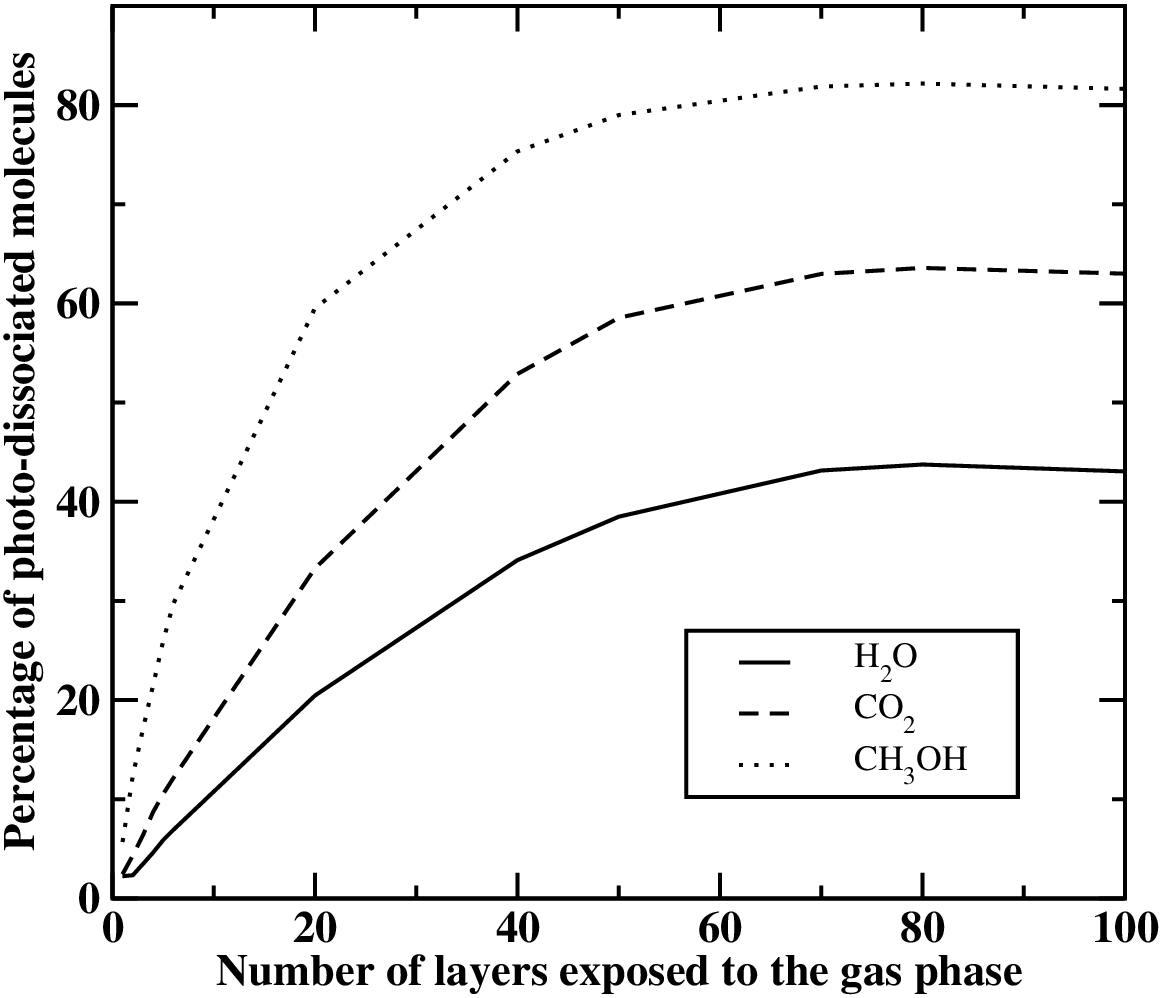}}
\caption{Effect of photo-dissociation on the number of layers exposed to the gas phase. Note that beyond fifty
mono-layers, there is a saturation in photo-dissociation.}
\label{fig-3}
\end {figure}
It is clear from the figure that in the Monte Carlo method,
the water (1\%-98\%), methanol (6\%-95\%) and CO$_2$ (91\%-99\%) 
production is heavily influenced due to interstellar photons and the
effects are prominent. Photo-dissociations of CO ($0.07\%$-97$\%$) and OH 
($0.009\% - 42\%$) are significant in the Monte Carlo method at the lower extinction region.
Around A$_V=3$, the percentage of photo-dissociation of CO and OH are $2.3\%$ and $0.19\%$ 
respectively which turns into $97\%$ and $42\%$ respectively when $A_v=0.5$. This demonstrates
how strongly the grain surface species are affected by the interstellar UV radiation 
field around the low extinction region.

In the Rate equation method, the trend is more or less similar. However,  it is to be noted that  
the stable species are photo-dissociated in the Rate equation method more strongly 
as compared to the Monte Carlo method. OH and CO are comparatively less affected.
The reason behind the enhanced photo-dissociation effect for the stable species 
in case of the Rate equation method is that this method is not capable of tracing the species layer wise
or location wise. Hence the molecules are dissociated with a constant fraction always. However, for the
Monte Carlo method, since we are considering the photo effect responsible for the first layer only, 
its contribution is lower than the Rate equation method. In the case of Rate equation 
method, trace amount of OH or CO is left over due to the formation of water but in case of Monte Carlo
method, some OH may be sustained due to not having the suitable reactant partner or due to the `blocking effect'
For instance, if a reactive species is surrounded by various surface species with which
reactions are not allowed or require high activation barrier energies, the species remains locked 
in until its desorption. If it overcomes the energy barrier during its lifetime on 
the grain or by any other mechanism such as evaporation or hopping of the surrounding 
species to free its location, then only the reactive species is unblocked. This is the 
`blocking effect' and it is a very common feature around the dense cloud, 
where the surface species are very much abundant. 
Due to this reason, in the Monte Carlo method there are some OH or CO
which are photo-dissociated as they remain in the grain for some time. 
On the contrary, in the case of Rate equation method this is not possible due to the
production of molecules at a constant rate.
\begin {figure}
\vskip 2.0cm
\centering{
\includegraphics[width=7cm]{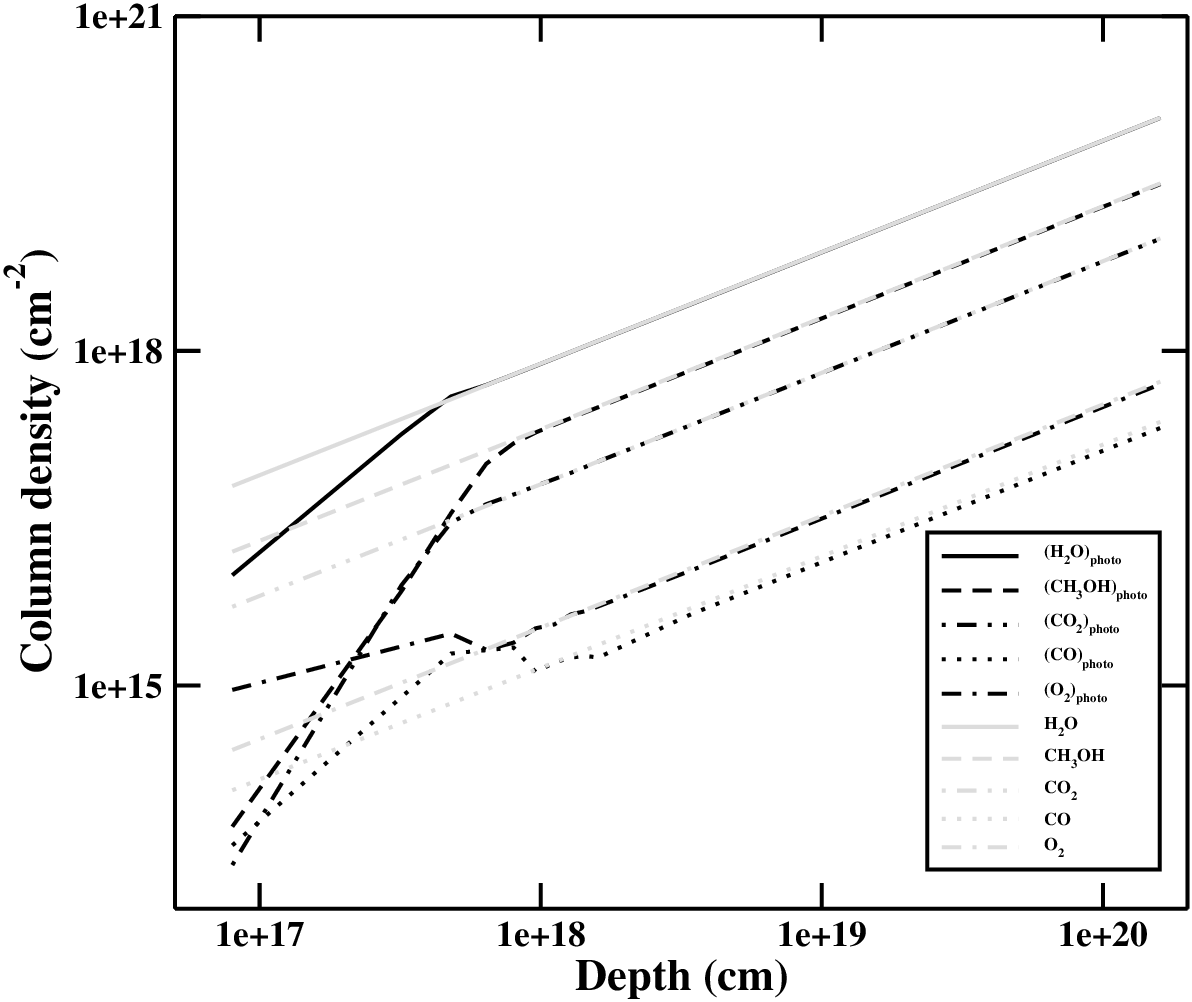}}
\caption{Column density of various species as a function of the distance inside the cloud 
measured from the surface. A comparison between the consideration of with and without
the photo effects is shown.}
\label{fig-3}
\end {figure}

\subsubsection{Effects of the choice of the exposed layers}

Interstellar radiation field is attenuated by the dust particles and thus the effects 
are also reduced with depth. The incoming particles have the 
energies in the range of $20-70$MeV nucleon$^{-1}$ which can deposit energy 
up to $0.4$ MeV \citep{Leger85} on an average into the 
classical size dust particles having radius $\sim 0.1$ $\mu$m. 
Several models are present to explain the interaction of interstellar energetic particles with the dust. 
Depending on the energy of the interacting particle, it can penetrate deep inside the grain mantle.
%But it has been observed that the effect of the energetic particles are mainly concentrated towards the topmost
%layers of the grain mantle. 
Recent chemical model by \citet{And06} considered photo-dissociation of 
the top six layers only. \citet{Ngu02} assumed a limit of one hundred mono-layers that could be 
reached by photo-dissociation.
\citet{West95} assumed that most of the desorption takes place only from the outermost mono-layers.
In Fig. 2, we have explained how the final results are altered by the assumption of the number of
exposed layers to the gas phase. {\it  In all the cases Photo-desorption is asumed to be effective 
only from the top most layer and photo-dissociation is assumed to be effective upon 
the choice of the exposed layers.}
Along X-axis, we plot the choice of the number of upper layers 
which can be affected by the inter-stellar photon. Along Y-axis, we plot the percentage
of photo-dissociated species at the end of the simulation time ($2 \times 10^6$ year). Percentage is
calculated by just calculating the ratio of total number of photo-dissociated H$_2$O, CO$_2$ or CH$_3$OH 
molecules with the total number of H$_2$O, CO$_2$ or CH$_3$OH produced during the simulation and
multiplying the ratio by $100$.
The life time of a molecular cloud is in general few million years. We restricted our simulation time
up to 2 $\times$ 10$^6$ year. In Fig. 2, results are presented for the intermediate density (10$^4$ cm$^{-3}$)
cloud after 2 $\times 10^6$ year, a steady state of the grain surface species is reached due to the depletion of the
gas phase species by that time. As the gas phase composition changes, production of stable molecules are becoming less
significant on the grain but the photodissociation rate is not changing. As a results, percentage of photo-dissociated
molecules are dependent on the simulation time. But, since we are presenting the results at the end of the 
life time of the molecular cloud, in that sense results showing the final percentage of the photo-dissociated molecules.
 
As expected, the percentage of photo-dissociated molecules 
increases as the number of exposed layers increases. Mantle thickness grows up to $\sim 80-85$ layers 
for the physical conditions considered here. This calculation is carried out assuming the DAC10
parameters as the initial condition. Results are plotted after evolving the system up to two million years.
After sixty layers the effects are slowed down.
For the higher density region, where mantle grows up to several mono-layers these effects should be 
more enhanced. But one point should be remembered that in this case, the field strength itself will be attenuated, 
as it go deep inside the mantle and the effects might be visible for the top-most mono-layers only.

In our work, we have tested how the choices of number of exposed layers may affect the final results.
Because the mantle column density is very low, 
we assumed that all the exposed layers, considered in the respective simulations, may `feel'
the effects of interstellar photons in a similar manner. In principle, it should have been decreased as it is going 
deep inside the mantle, but this may be negligible.
%For example, if a grain mantle has $n$ number of mono-layers,
%the top most layer and $n-1$ layers should not have the same effect. It must be reduced by some extent.
%But this factor is yet to be known with certainty. Our results shows, strong variations with the choice 
%of number of exposed layers. This should not be the case if we know the correct penetration 
%factor and that factor would then decide after how many layers, contribution of photo-process are negligible.
In the rest of the paper, we consider two extreme cases:
When the photo-dissociation is effective only for (a) the outermost layer and (b) uppermost fifty layers.

\subsubsection{Column density}
When we study molecular abundances, we do not see any local molecular density, but we observe only the column 
densities of the molecules. So theoretical calculation of the column density of a given molecule should be done
in order to compare the theoretical and observational results. The column density 
of a species is given by (\citet{Shala94}),
\begin{equation}
N(A)=\Sigma n_i(A,R_i) \Delta R_i,
\end{equation}
where, n$_i$(A,R$_i$) is the local number density of species $A$ at location $R_i$ 
and $\Delta R_i=R_i-R_{i-1}$ is the grid spacing along the radial($R$) direction. Number density 
of the species at the location $R_i$ is given by, $x_i=\frac{n_i(A,R_i)}{{n_H}_i}$,  
where ${n_H}_i$ is the number density of the cloud at the $i^{th}$ location.
Following \citet{Shala94}, length of the line of sight R$_{tot}$ can be represented as;
\begin{equation}
R_{tot}=\frac{1.6 \times 10^{21} \times A_v}{{n_H}_i} \ {\rm cm} ,
\end{equation}
where, A$_V$ is the visual extinction of cloud.
Taking summation over the entire grid (total $k$ number of grids),
\begin{equation}
\Sigma_1^k x_i= \Sigma_1^k \frac{ n_i(A,R_i)}{{n_H}_i} .
\end{equation}
So the column density of a species `A' is given by,
\begin{equation}
N(A)=\Sigma_i^k {n_H}_i x_i \Delta R_i .
\end{equation}
For the constant density cloud, ${n_H}_i={n_H}$. We get,
\begin{equation}
N(A)=n_H \Sigma_i^k x_i \Delta R_i.
\end{equation}
For $k=1$;
\begin{equation}
N(A)=n_H x_i \Delta R_i.
\end{equation}
Thus, by knowing the extinction parameter, we can calculate the depth of a cloud from Eqn. 5 and the 
corresponding Column density from Eqn. 9. In Fig. 3, we show the depth dependence of the column density 
of the major interstellar grain surface species with and without interstellar photo effects. We 
assumed the initial conditions as in DAC10. From Fig. 3, it is clear that the 
column density increases with the depth of the cloud (measured from the outer edge), 
which indicates that as we are going deeper inside the cloud, production enhances. Pictures should 
be better interpreted if we assume that the extinction parameter consists of two components: the first one is $\le$ 5 
(corresponding depth $8 \times 10^{17}$ cm, from Eqn. 9) and the next one is $\geq$ 5. For this figure, 
we have converted the final abundances into the column density after $2 \times 10^6$ years. Low value 
of the extinction parameter means diffuse or weakly translucent cloud, so it is likely to evolve 
the cloud with much longer time (around $10^7$ years, here we evolve the cloud only up to
$2 \times 10^6$ year). In this lower region, we find out a very stiff slope of the
column density with the extinction parameter/depth of the cloud, whereas at the moderate translucent cloud region 
($A_v \ge 5$) we have almost a linear relationship. For the lower end of the extinction parameter or equivalently
when length of the line of sight is smaller, we still have a significant water formation during the life time
($2 \times 10^6$ year) of the molecular cloud. 
According to \citet{Mur00}, the observed column density of the 
condensed water molecules along the lower edge of the extinction parameter {$\sim$ 0.5} is around $10^{17}$  
cm $^{-2}$.  \citet{Mur00} studied the Heiles Cloud 2, which is a part of the Taurus
molecular cloud complex. Average number density of the dense core is around 10$^4$ cm$^{-3}$ \citep{Onish96}.
From Fig. 3, our calculated column density of condensed phase water is around $1.1 \times 10^{16}$ 
cm$^{-2}$. For the lower extinction region, cloud may have larger life time,
which means that we need to calculate the column density after much longer time than the previous cases.
For the sake of completeness, we tested a case for $A_v=0.5$ (corresponds to the depth of $8 \times 10^{16}$ cm)
for 10$^7$ years and obtain the column density of water ice around $4.6 \times$ 10$^{16}$cm$^{-2}$. 
This is much better than the previous estimate.
Photo effect can be better understood if we make a comparison 
of two cases. In one case we include this effect, and in the other case
we exclude this effect. In Fig. 3, gray lines are drawn for the cases 
where we have not considered the photo effects. Difference between the
two results are distinctly visible towards the low extinction region.
For the better realization of these differences, 
we have tabulated the column densities of the different species at Table 4
for A$_V=0.5$ (corresponding depth 8$\times 10^{16}$ cm), which is the lowest value of the 
extinction used in this present paper. It is clear from the Table 4 that in the region of very low extinction, 
composition of the grain mantle is completely different due to the interstellar photo effect. Column densities
of water \& methanol are strongly affected by the interstellar photo
reactions. Column densities of all the species are decreased except O$_2$. Interestingly, it is increasing
around these region. The reason behind this increased column density of O$_2$ is that it can be 
easily photo-dissociated by forming two Oxygen atoms, which in turn react to form O$_2$ molecules again. 
Source of atomic oxygen may also come from the  photo-dissociation of OH, CO$_2$, HO$_2$, O$_3$ and CO. Due to the
large number of O atoms on the grain surfaces, O$_2$ abundances increased (Fig. 3).
\begin{table}
\caption{Effect of photo-dissociation for A$_V=0.5$}
\begin{tabular}{ccc}
\hline
Species&No photo effect& With photo effect\\
\hline\hline
H$_2$O&6.1 $\times 10^{16}$&1.1$\times 10^{16}$\\
CH$_3$OH&1.6 $\times 10^{16}$&5.4$\times10^{13}$\\
CO$_2$&5.1 $\times 10^{15}$&3.3$\times 10^{14}$\\
CO&1.2$\times 10^{14}$&2.1$\times 10^{13}$\\
O$_2$&2.7 $\times 10^{14}$&1$\times 10^{15}$\\
\hline
\end{tabular}
\end{table}

\subsubsection{Relation between $A_v$ and hydrogen number density}

%So far, we have assumed that the extinction parameter does not depend on other physical parameters. 
%However, in principle, we need to consider the extinction parameter by considering the appropriate physical parameters.
Following \citet{Lee96}, the relationship between hydrogen density and visual extinction is given by,
\begin{equation}
n_H={n_H}_0 [1+[{(\frac{{n_H}_{max}}{{n_H}_0})}^{1/2}-1)\frac{A_v}{{A_v}_{max}}]]^2,
\end{equation} 
where, $n_H=n(H)+2n(H_2$) is the total hydrogen density in the unit of cm$^{-3}$, n${_H}_0$ is the cloud
density at the cloud surface, n{$_H$}$_{max}$ is the maximum density and A$_V{_{max}}$ is the maximum visual extinction
considered very deep inside the cloud. \citet{Lee96} considered the above relations for $T=30$K and here for the 
limitation of our present model we assume that this relation is valid for $10$K also. We assume that at the lowest depth 
(i.e., at the cloud surface) hydrogen density $n{_H}_0$ is $10^3 \ cm^{-3}$ and it reaches its maximum $n{_H}_{max}=
10^5 \ cm^{-3}$ at deep inside the cloud (corresponding $A{_V}_{max}=150$). Composition of the gas phase around 
different regions are taken accordingly by just appropriate scaling of the DAC10 parameters. In Fig. 4, variation of column density
with respect to the extinction parameters are shown. Column density of the species at different interstellar extinction
is taken from results obtained at $2$ million years. The inset shows the variation of the number density for the
varying extinction parameters. As in Fig. 3, here also H$_2$O is the major constituent of the grain mantle.
Methanol is the $2^{nd}$ abundant contributor initially but deep inside the cloud its contribution slightly decreases.
The reason behind is that due to the increasing number density around the region of high extinction parameters, 
accretion rates of O and CO increase and as a result, the surface coverage of O related 
species increases (CO$_2$ and O$_2$ increases). The accretion rate of atomic hydrogen 
is not increasing as we assumed that it is constant ($1.10$ cm$^{-3}$) around the all region, though in principle, it should 
have decreased slightly around the high density region. Due to the lack of atomic hydrogen on the grain surfaces, 
all the hydrogenated species are looking to be produced less efficiently around the high density/high extinction region. As a 
result, the surface coverage of CO increases rapidly around these regions.
\begin {figure}
\vskip 2.0cm
\centering{
\includegraphics[width=7cm]{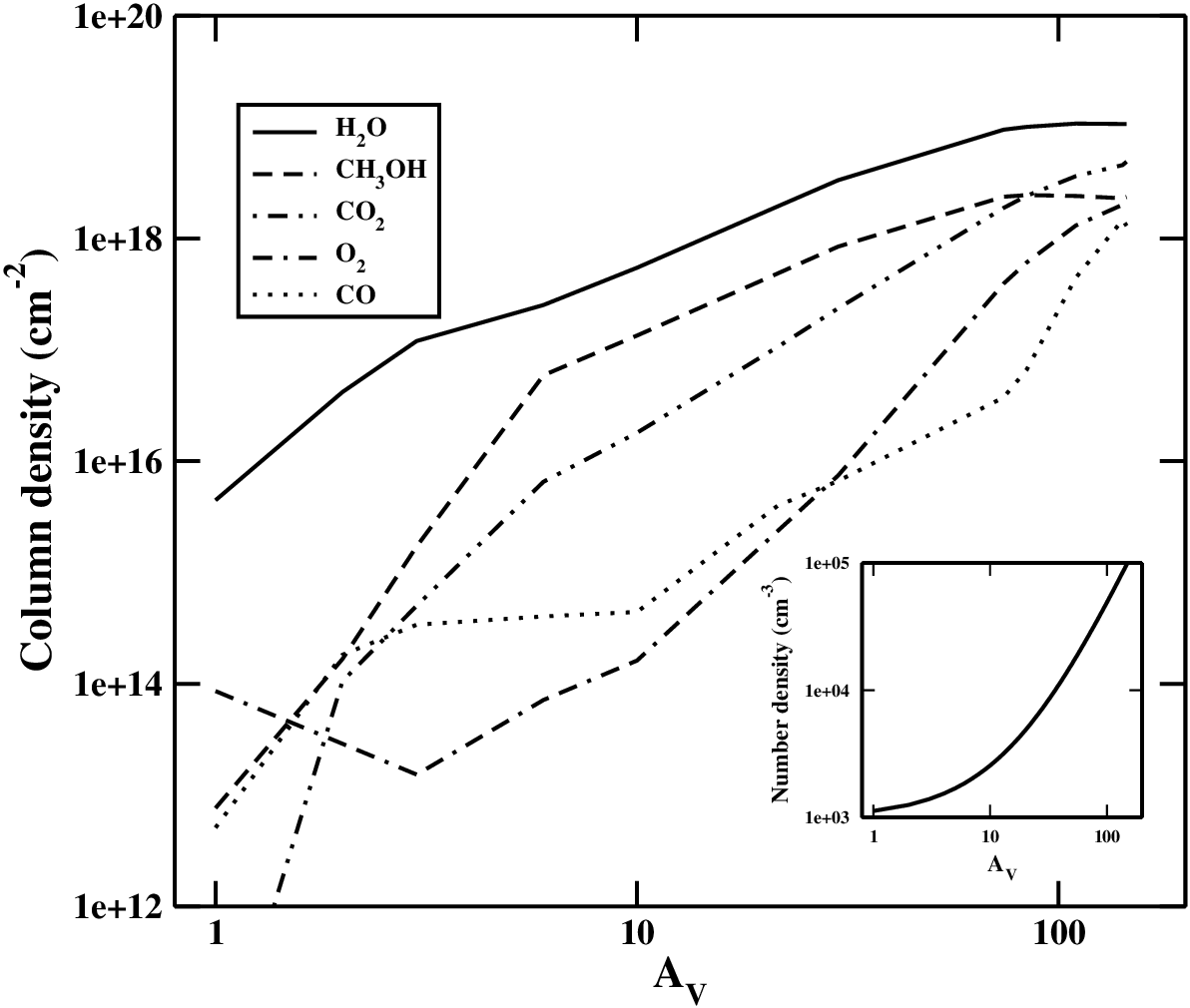}}
\caption{Variation of the column density with the extinction parameter. Inset shows the variation of density with the
change of the extinction parameter.}
\label{fig-1}
\end {figure}

\subsubsection{Time dependent Case}

Armed with various parameters, we will now simulate more realistic case, where 
the density is not constant with time. To start with, we assume that a cloud is collapsing from 
a low density region ($n{_H}_0=10^3$ \ cm$^{-3}$) to the higher density ($n{_H}_{max}=$10$^5 \ cm^{-3}$) 
region. Density is assumed to be increasing linearly with time and reaches its maximum
at around $2$ million years. We assume that around the high density region, the cloud has 
the maximum value of the extinction parameter (${A_v}_{max}=150$) and the cloud 
is assumed to be collapsing isothermally ($T=10$K). Using Eqn. 10, we can calculate
the corresponding extinction parameter. So in that way we now have a pseudo dynamic 
nature of the interstellar collapsing cloud where we have a relation 
between the time and the other parameters. The gas phase composition changes 
according to the density of the surrounding molecular cloud. The initial compositions 
are taken by appropriate scaling of the DAC10 parameters and effects of interstellar photons are
considered only for the uppermost layer.
Interaction of the gas 
and the grain are considered by assuming the accretion and evaporation process. 
Fig. 5 shows the time variations of column densities. Column density of species
at a particular instant is calculated by just converting the abundance by Eqn. 9. It is evident from the figure that
at the beginning, since the density is pretty low and the extinction is much lower, the
longevity of the several species are restricted due to the strong radiation field. 
Inset pictures show the variation of the (a) hydrogen number density and 
the (b) extinction parameter with the time. At the beginning, due to the strong UV radiation 
field (low extinction parameter), the column density of several species fluctuates. After 10$^5$ 
years, variations are more or less smoother due to the decrease in the radiation field in the high extinction parameter
region. As like the other cases, here also water is the most abundant species. Surface coverage
of methanol is noticed to be reduced heavily in compare to the other results.
\begin {figure}
\vskip 2.0cm
\centering{
\includegraphics[width=7cm]{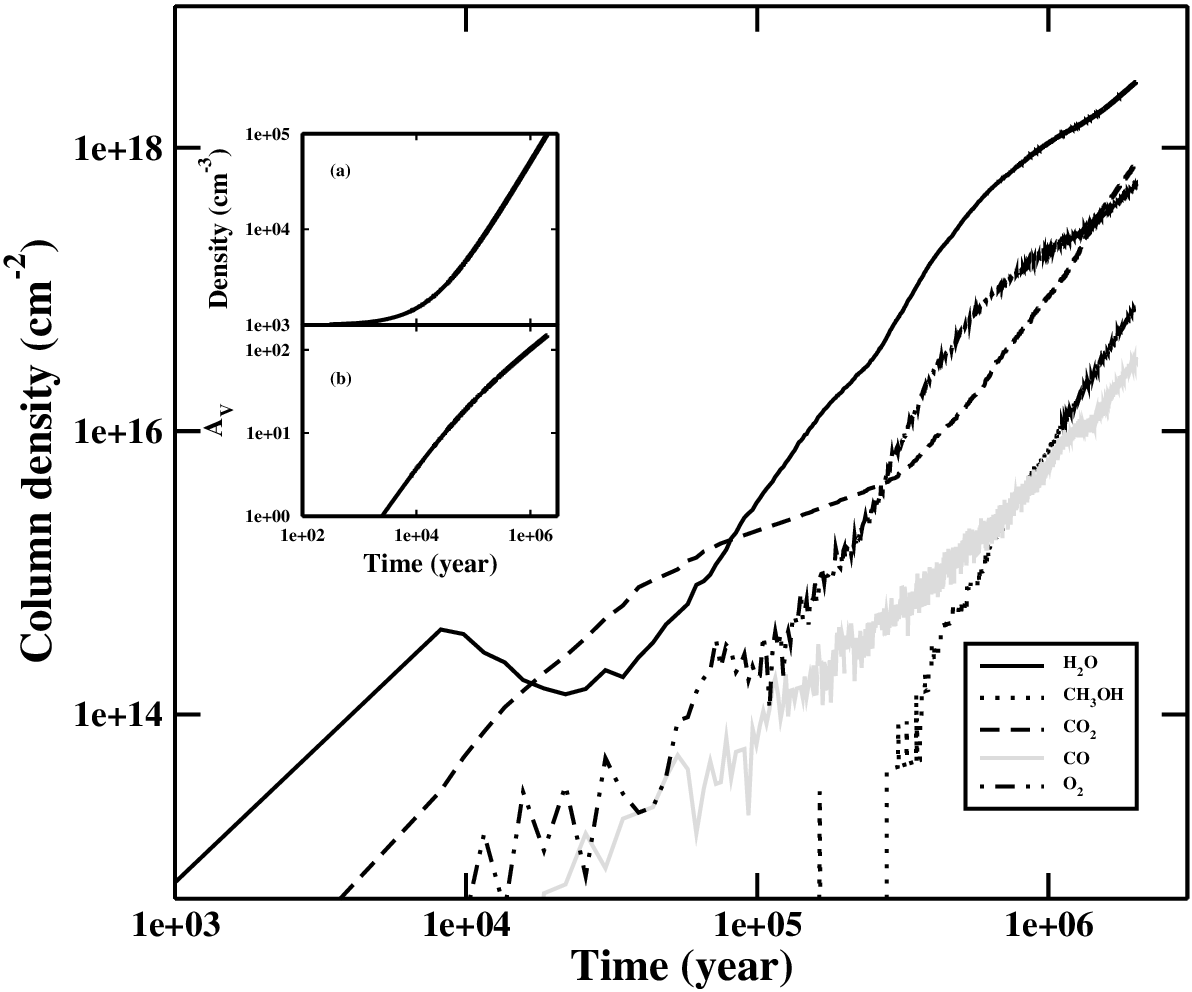}}
\caption{Time evolution of several condensed phase species are shown. Insets show the
(a) time variation of density and the (b) time variation of the extinction parameter.}
\end {figure}
\begin {figure}
%\vskip 2.0cm
\centering{
\includegraphics[width=7cm]{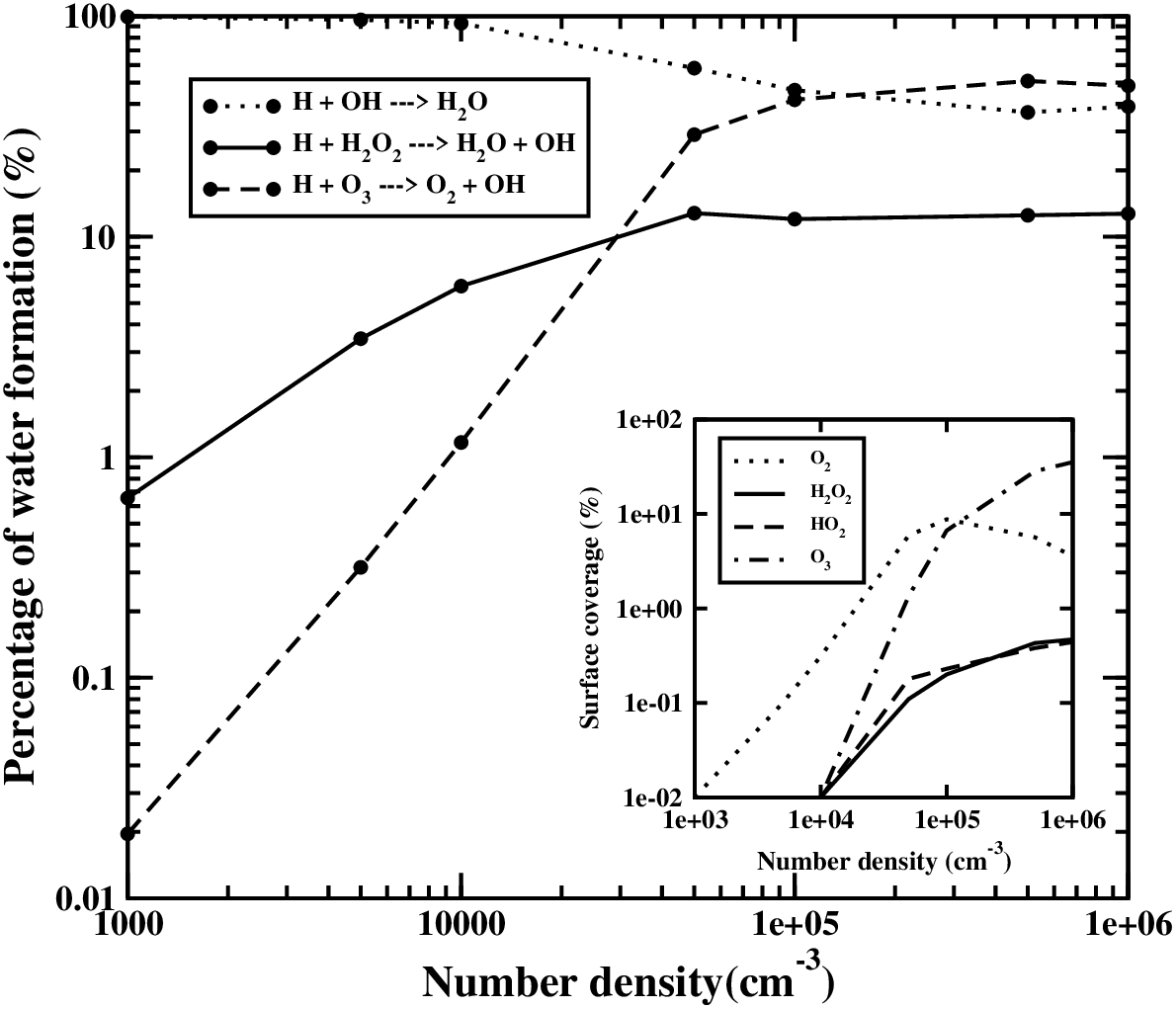}}
\caption{Different routes to H$_2$O.}
\end {figure}
\subsection{Different routes to water formation}
Recent experiment by \citet{Ioppolo08} shows that there are alternative routes
for the formation of water on the interstellar grain other than the traditional one (H+OH $\rightarrow$ H$_2$O). 
We have visited the problem of the water production via different routes by using our modified Monte Carlo code.
Fig. 6 shows that the water molecules are mainly formed by the well known traditional route
but as the density of the molecular cloud changes (for simplicity, we keep other physical constraints as constant: 
$A_v=10$ and $T=10$K), other pathways for the formation of water molecules also appears to be significant. 
For example, at lower density regions ($n\sim 10^3 cm^{-3}$), 
almost 100\% of the water molecules are formed by the traditional route but when the  
number density of hydrogen is around $10^6 \ cm^{-3}$, this goes down to 39\%.
At higher number density regions, production of O$_2$ is favorable and due to the high surface coverage
of solid O$_2$, O$_3$ can be produced very efficiently. Since
the activation barrier energy for the reaction between H and O$_3$ is very low ($450$K),
O$_3$ can readily react with H to form an OH, which in turn, form H$_2$O by another hydrogenation reaction.
Formation of H$_2$O by the O$_3$ route is very efficient at high density regions
(maximum $\sim 48\%$) compare to the low density region (around $\sim0.02\%$). There is another route
to H$_2$O, namely from H$_2$O$_2$. Reaction Nos. 12 to 14 are associated with the
formation of H$_2$O via this route. Formation is maximum around the high density region (13\%).
At around $n \sim 10^6$ cm$^{-3}$, production of O$_3$ is highly favorable due to
high surface coverage of atomic oxygen as well as O$_2$ molecules. O$_3$ molecules are produced efficiently
via Langmuir Hinshelwood or via Eley Rideal method, as a result, the surface coverage of O$_2$ decreases.
The inset clearly displays the above aspects
showing the variation of the surface coverage of several condensed phase species with the
number density of the molecular cloud. It is clear that the surface coverage of O$_2$ having a peak at around
10$^5$ cm$^{-3}$ region and beyond this region as the density increases, rate of production of O$_3$ enhances.
As a result, the surface coverage of O$_2$ decreases.
 
Around the dense region of the molecular cloud, H$_2$ is the most abundant species in the gas phase and
H$_2$O can be very efficiently produced in the gas phase via reaction number 16 of Tab. 1. 
But the problem with H$_2$ is that it can not able to stick well with the
grain surface \citep{Leitch85}. So we use sticking coefficient of H$_2$ to be zero.
As like the stantcheva 2002, Das et al., 2008b, and Das et al., 2010, here also we consider
only the accretion of H, O and CO from the gas phase.
In our model, we allow gas phase H$_2$ to react with surface OH via Eley-Rideal mechanism. But, as
H$_2$ + OH reaction requires very high activation energy (2600K) \citet{Schiff73} and
surface abundance of OH is very low, production of water via this route is negligible.
H$_2$ may produce on the grain surface via reaction between two H atom and can 
contribute to the water formation route. But if we look at the accretion time scale of 
H atom onto a 50$\times$50 grain kept at T$\sim$10K, it comes out to be around $4.8\times10^7$ sec 
for 10$^4$ cm$^{-3}$ number density cloud. The evaporation  time scale at 10K
is around 535 sec. So there is no chance for one H to meet with other H via accretion.
Despite of this fact H$_2$ may produce on the grain surface when two H atom meet together. For that instant
one H is the accreted H atom and the second one is the product due to the reaction between O and
HCO (reaction number 10 of Tab.1). Since the production of H$_2$ by this way also not significant enough, 
this reaction does not significantly contributes to the water formation on the grain.

\subsection{Comparison with Observations and other Models}
NGC 7538 IRS 9, W33A, W3 IRS 5, S140 IRS 1, RAFGL 7009S were chosen to compare our results 
with observations. The reasons behind choosing these sources are that these sources are believed to
be protostellar objects which are deeply embedded inside dense molecular clouds.
Beside this, they have a very high level of extinction from dust in the 
dense molecular cloud and they exhibit infrared spectra with the molecular ice feature. 
In Table 5, observed column densities as well as relative abundances of abundant observed ice species 
(H$_2$O, CH$_3$OH, CO$_2$, CO and O$_2$)
along different lines of sight are compared with our various modeling results.
Column densities are expressed in the unit of 10$^{17}$ cm$^{-2}$. 
Normally relative abundances of ice species are used to express the observed ice features
and this is calculated by just calculating the surface abundance of any species with respect 
to the surface abundance of water and then multiplying the factor by 100.
Observed relative abundances along various sources as well as the relative abundances obtained from 
our simulation results are noted inside the bracket.
Appropriate references for the observed data are noted at the bottom of the table.
For the comparison purpose, four models are considered; DAC10a, DAC10b, DAC10a(m) and DAC10b(m). 
In DAC10a and DAC10b model, initial conditions are taken around the favourable zone of the
parameter space \citep{Das10}, having number density of the cloud 10$^4$ cm$^{-3}$ and temperature 10K.
Difference between DAC10a and DAC10b model is that, for DAC10a
model, effect of photo reactions/evaporations are considered only for the outermost layer and for
DAC10b model, Photo-dissociation is considered for the uppermost 50 layers but photo-evaporation is 
restricted only from the topmost layer. 
In DAC10a and DAC10b model, extinction parameter is assumed to be independent upon the initial number density 
of the cloud. In DAC10a(m) and DAC10b(m) model, density variation of the 
extinction parameters are taken into account (Eqn. 10). 
In the inset picture of Fig. 4, variation of number density with respect to the  
extinction parameter of the cloud is shown.
These models may be better understood by just referring the section 3.1.4. Difference between DAC10a(m) and
DAC10b(m) model is the same as the difference between the DAC10a and DAC10b model mentioned just above.
Column densities from our simulation results are tabulated after two million years. Column density is
calculated by using Eqn. 9,  where column density is directly proportional to the extinction parameter (A$_V$). 
While we are considering the effect of photo-reactions for the uppermost 
50 layers (DAC10b and DAC10b(m) model), production of
the stable species(H$_2$O, CO$_2$, CH$_3$OH and O$_2$) decreases for most of the cases compare to the cases, 
where only topmost layer is considered for the effective photo reactions/evaporations (DAC10a and DAC10a(m) model).
Opposite trend is noticed for the CO molecule. Since in case of DAC10b and DAC10b(m) model, a large number of CO$_2$
molecules can be affected by photodissociations (CO$_2$ $\rightarrow$ CO +O) compare to the DAC10a and DAC10a(m) 
model respectively, surface coverage of CO molecule increases.

In Fig. 3,  DAC10a model was used, where number density of the cloud is assumed to be 
independent upon the extinction parameter.
It is clear from Fig. 3 that beyond A$_V$=10, this effect is negligible. 
We have carried out similar simulations for the DAC10b model also and noticed the 
similar trend. In Table 5, all the observed sources having extinction 
parameter beyond 10. Due to this reason, in Table 5, relative concentration of the ice
species along different lines of sight are constant for DAC10a and DAC10b model. 
Column density for all the models varies, since it is proportional to the extinction parameter by Eqn. 9.
In case of the DAC10a(m) and DAC10b(m) model, initial number density is totally different 
compare to DAC10a and DAC10b model (Eqn. 10) and as a results, in Table 5, relative concentration
along different lines of sight also differs.

There are several chemical models for explaining the chemical evolution 
of the interstellar grain mantle by Monte Carlo method.
Our Monte Carlo model is distinctly different from the other Monte Carlo models used by several authors 
\citep{Cazau10, Cup07, Cup09, Chang05}. 
Our initial gas phase composition is also different in compare to the other models.
We use initial composition of gas phase according to the favourable zone mentioned in \citet{Das10}. 
\citet{Chang05} studied the formation of molecular hydrogen only on the grain surfaces
by Monte Carlo procedure so it is not possible to compare our results directly with them.
Main difference between ours and \citet{Cazau10}, 
is that they did Monte Carlo simulation to follow the chemistry involving H, D, and O. 
They did not consider the chemistry involving CO.
In our case, we consider the chemistry between H, O and CO on the interstellar grain and
Deuterium chemistry yet to be considered in our model. 
\citet{Cup07}, studied the formation of water on grain surface for diffuse,
translucent and dense clouds. They did not consider the formation of CH$_3$OH and CO$_2$
on the grain. As a results, grain mantle was always dominated by the water. In \citet{Cup09}
formation of CH$_3$OH was considered and it is noticed that in our case 
H$_2$CO is under produced, in comparison with the \citet{Cup09}.

\begin{table*}
\begin{center}
\caption{\bf Column densities of the ice species in units of 10$^{17}$ cm$^{-2}$ 
and relative abundances of ice species(inside the bracket) along different
lines of sight}
\begin{tabular}{ccccccc}
\hline
{\bf Species}&&{\bf W33A}&{\bf W3 IRS 5}&{\bf NGC 7358 IRS 9}&{\bf S 140 IRS 1}&{\bf RAFGL 7009S}\\
&&(\bf A$_V$=145)&(\bf A$_V$=142)&(\bf A$_V$=84)&(\bf A$_V$=74.1)&(\bf A$_V$=21.6)\\
\hline
&\bf{Observed}&90-420$^a$(100)&43-58$^a$(100)&70-110$^a$(100)&21-88$^a$(100)&110$^c$(100)\\
{\bf H$_2$O}&\bf{DAC10a}&177.5(100)&173.8(100)&102.8(100)&90.7(100)&26.4(100)\\
&\bf{DAC10b}&108.4(100)&106.2(100)&62.8(100)&55.4(100)&16.2(100)\\
&\bf{DAC10a(m)}&106.7(100)&106.9(100)&100.6(100)&94.8(100)&19.5(100)\\
&\bf{DAC10b(m)}&71.4(100)&72.1(100)&66.5(100)&61.6(100)&12.2(100)\\
\hline 
&\bf{Observed}&39-230$^a$(5$^h$)&5.3-35$^a$(8.4$^h$)&9.1-67$^a$(3.2$^h$)&3.8$^a$(6.8$^h$)&33-38$^f$(30$^h$)\\
{\bf CH$_3$OH}&\bf{DAC10a}&45.3(25.5)&44.4(25.5)&26.2(25.5)&23.2(25.5)&6.8(25.5)\\
&\bf{DAC10b}&29.8(27.5)&29.2(27.5)&17.3(27.5)&15.2(27.5)&4.4(27.5)\\
&\bf{DAC10a(m)}&23.6(22.1)&23(21.5)&24.6(24.4)&23.7(25)&4.8(25)\\
&\bf{DAC10b(m)}&24.6(34.4)&24.4(33.9)&19.8(29.7)&18.4(29.7)&2.8(22.8)\\
\hline
&\bf{Observed}&14.5$^e$(3.6$^h$)&7.1$^e$(11.3$^h$)&4.6-7.3$^d$(16.3$^h$)&4.2$^e$(7.5$^h$)&0.4-25$^c$(21$^h$)\\
{\bf CO$_2$}&\bf{DAC10a}&14.7(8.3)&14.4(8.3)&8.5(8.3)&7.5(8.3)&2.2(8.3)\\
&\bf{DAC10b}&7.0(6.4)&6.8(6.4)&4.0(6.4)&3.6(6.4)&1.0(6.4)\\
&\bf{DAC10a(m)}&48.9(45.8)&45.7(42.8)&24.2(24)&18.8(19.9)&1.1(5.5)\\
&\bf{DAC10b(m)}&26.1(36.6)&24.5(34)&14.4(21.7)&11.2(18.2)&0.5(4.1)\\
\hline
&\bf{Observed}&2.8$^a$(2.2$^h$)&0.54-1.1$^a$(2.5$^h$)&3.2-6.4$^a$(12$^h$)&-(0.4$^h$)&18$^c$(15$^h$)\\
{\bf CO}&\bf{DAC10a}&0.29(0.17)&0.29(0.17)&0.17(0.17)&0.15(0.17)&0.04(0.17)\\
&\bf{DAC10b}&3.0(2.7)&2.9(2.7)&1.7(2.7)&1.5(2.7)&0.44(2.7)\\
&\bf{DAC10a(m)}&13(12.19)&14.7(13.75)&0.66(0.66)&0.37(0.39)&0.04(0.2)\\
&\bf{DAC10b(m)}&29.7(41.6)&29.6(41.1)&6.5(9.7)&4.2(6.7)&0.24(2.0)\\
\hline
&\bf{Observed}&-(-)&-(-)&12$\pm$5$^g$(-)&-(-)&-(-)\\
{\bf O$_2$}&\bf{DAC10a}&0.73(0.41)&0.71(0.41)&0.42(0.41)&0.37(0.41)&0.11(0.41)\\
&\bf{DAC10b}&2.9(2.64)&2.8(2.64)&1.7(2.64)&1.5(2.64)&0.43(2.64)\\
&\bf{DAC10a(m)}&20.2(18.9)&20.1(18.8)&6.1(6)&3.9(4.1)&0.02(0.12)\\
&\bf{DAC10b(m)}&9.3(13.1)&9.4(13.1)&3.7(5.5)&2.7(4.4)&0.28(2.3)\\
\hline
\end{tabular}
\end{center}
\begin{flushleft}
\scriptsize{
 $^a$ \citet{Alla92}\\
            $^b$ \citet{Jiang00}\\
            $^c$ \citet{D'Hend96}\\
            $^d$ \citet{Whit96}\\
            $^e$ \citet{Gibb04} \\
            $^f$ \citet{Dart99}\\
            $^g$ Upper limit derived by \citet{Vand99}}\\
            $^h$ \citet{Kean01}\\
\end{flushleft}
\end{table*}

\section{Conclusions}

In this paper, we carried out a Monte Carlo simulation to study the
formation and structural information of the interstellar grain mantle.
The main results of our study are as follows:

\noindent $\bullet$ The effects of interstellar radiations are
included to extract the information about the structure of the
interstellar grain mantle. The effects of photo-dissociation are
studied as a function of the extinction parameter. For lower
values of the extinction parameter, the photo-dissociation effects
dominate. A comparison between the Rate equation method and the Monte
Carlo method has been made and difference in results between the two
approaches is presented in Fig.1. We notice that in comparison with the Monte
Carlo method, the Rate equation method always
overestimates the abundances of species.
Stable species like H$_2$O, CH$_3$OH and
CO$_2$ are photo-dissociating more strongly in the Rate equation
method as compared to the Monte Carlo method.

\noindent $\bullet$ Recent chemical models assumed that the
interstellar photons can have effects on the first few mono-layers
only. But as the column density of the grain mantle itself is very
low, we assume that the effect may be significant also deep inside the
mantle. We notice that the choices of number of exposed layers can
affect the chemical composition of the interstellar grain mantle
significantly.

\noindent $\bullet$ Column densities are calculated along different
lines of sight and it is noticed that deep inside the cloud, the
probability of finding different grain surface species are very high.
Effects of interstellar photons are extensively studied and it is
noticed that beyond $A_V$=10, its contribution toward the chemical
evolution is negligible.

\noindent $\bullet$ Column densities of various grain surface species
are studied for static as well as time dependent models. A static
model is considered which is more realistic than the 
DAC10 model, where the number density of the cloud is 
varied according to the extinction parameter of the cloud. 
As the initial conditions are completely different
from those of the DAC10 model, our results are also significantly
different. In the time dependent model, variation of the
number density and extinction parameter with time is considered to
follow the chemical evolution during the collapsing phase of a
proto-star. It is noticed that since initially the density
was low, formation of several species was seriously hindered by the
effects of the interstellar photon. As the time passes by, the density and
A$_V$ increases, which in turn enhance the production of several
grain surface species.

\noindent $\bullet$ Water is the most abundant species on the grain.
There are several routes by which water can be formed on the
interstellar grain. Major parts of H$_2$O are formed via traditional
route(H + OH). But there are also other pathways responsible for the
H$_2$O production on the grain. From our simulation, we find that
the formation of H$_2$O  via O$_3$ and H$_2$O$_2$ routes are
also have significant contributions, especially in high density regions.

\noindent $\bullet$ A comparison between the results of our approaches
and observational data has been made along different lines of sight.
The initial gas phase composition
was chosen from the DAC10 model (DAC10a and DAC10b) as well as the
modified DAC10 model (i.e., by considering the variations of density
with the extinction parameter, DAC10a(m) and DAC10b(m)).
Some of our results are found to be in good agreement with different
observational results obtained so far.

\section{Acknowledgments}
This work was partly supported by a DST project (Grant No. SR/S2/HEP-40/2008).


\begin{thebibliography}{}
\bibitem[\protect\citeauthoryear{Allamandola et al.,}{1992}]{Alla92} Allamandola, L. J., 
Sandford S. A. \& Tielens A.G.G.M. 1992., APJ, 399, 134 
\bibitem[\protect\citeauthoryear{Allen \& Robinson}{1975}]{Allen75}Allen, M., Robinson, 
G. W., 1975., ApJ, 195, 81 
\bibitem[\protect\citeauthoryear{Allamandola et al.}{1988}]
{Alla88}Allamandola, L. J., Sandford, S. A., Valero, G. J. 1988, Icarus, 76, 225
\bibitem[\protect\citeauthoryear{Andersson et al.,}{2006}]{And06} 
Andersson, S., Al-halabi, A., Kroes, G., Van Dishoeck, E. F.,
2006, JCP, 124, 064715
\bibitem[\protect\citeauthoryear{Boogert \& Ehrenfreund}{2004}]{Boo04}
Boogert, A.C.A \& Ehrenfreund, P., 2004, ASPC 309, 547
\bibitem[\protect\citeauthoryear{Cazaux et al.,}{2010}]{Cazau10} 
Cazaux, S.,Cobut, V., Marseille, M., , Spaans, M., Caselli, P.,
2010, A\&A, 522, A74
\bibitem[\protect\citeauthoryear{Chang et  al.}{2005}]{Chang05}Chang, Q., Cuppen, H., M., Herbst, E., 
2005, A\&A, 434, 599
\bibitem[\protect\citeauthoryear{Charnley}{2001}]{Charn01}Charnley, S.B., 2001, ApJ, 562L, 99
\bibitem[\protect\citeauthoryear{Chakrabarti et al.}{2006a}]{Chak06a}Chakrabarti, S.K., Das, A., Acharyya, K.,
Chakrabarti, S., 2006, A\&A, 457, 167
\bibitem[\protect\citeauthoryear{Chakrabarti et al.}{2006b}]{Chak06b} Chakrabarti, S.K., Das, A., Acharyya, K.,
Chakrabarti, S., 2006, BASI, 34, 299
\bibitem[\protect\citeauthoryear{Cuppen \& Herbst}{2007}]{Cup07}Cuppen, H. M., Herbst, E., 2007, APJ, 668, 294
\bibitem[\protect\citeauthoryear{Cuppen et al.}{2009}]{Cup09}Cuppen, H. M.,
Van Dishoeck E., F., Herbst, E., 
Tielens, A. G. G. M., 2009, A\&A, 508, 275
\bibitem[\protect\citeauthoryear{Dartois et al.,}{1999}]{Dart99} 
Dartois, E., Schutte, W., Geballe, T. R., Demyk, K., Ehrenfreund, P., D'Hendecourt, L. 1999
A\&A, 342L, 32
\bibitem[\protect\citeauthoryear{Das et al.}{2008b}]{Das08b}Das, A., Acharyya, K., 
Chakrabarti, S., Chakrabarti, S. K., 2008, A \& A, 486, 209
\bibitem[\protect\citeauthoryear{Das et al.,}{2010}]{Das10} 
Das, A., Acharyya, K., Chakrabarti, S. K., 2010, MNRAS 409, 789
\bibitem[\protect\citeauthoryear{Das et al.}{2008a}]{Das08a}Das, A., Acharyya, K., 
Chakrabarti, S., Chakrabarti, S. K., 2008, New Astronomy, 13, 457
\bibitem[\protect\citeauthoryear{D'Hendecourt et al.,}{1996}]{D'Hend96} 
D'Hendecourt, L., et al., 1996, A\&A, 315, L365
\bibitem[\protect\citeauthoryear{D'Hendecourt al.}{1982}]{Dhen82}D’Hendecourt, L. B., 
Allamandola, L. J., Baas, F., Greenberg, J. M. 1982, A\&A, 109, L12
\bibitem[\protect\citeauthoryear{Draine et al.,}{2003}]{Draine03} 
Draine, B. T., 2003, APJ 598, 1017.
\bibitem[\protect\citeauthoryear{Fuchs et al.,}{2009}]{Fuchs09} 
Fuchs, G. W., Cuppen, H. M., Ioppolo, S., Romanzin, C., Bisschop, S. E.,
Andersson, S., van Dishoeck, E. F., Linnartz, H., 2009, A\&A, 505, 629
\bibitem[\protect\citeauthoryear{Goumans \& Andersson}{2010}]{Goumans10} 
Goumans, T., P., M., Andersson, S., 2010, MNRAS, 406, 2213.
\bibitem[\protect\citeauthoryear{Gibb et al.,}{2004}]{Gibb04} 
Gibb, E. L., Whittet, D. C. B., Boogert, A. C. A., Tielens,
A. G. G. M., 2004, ApJS 151, 35
\bibitem[\protect\citeauthoryear{Hagen et al.}{1979}]{Hagen79}Hagen, W., 
Allamandola, L. J., Greenberg, J. M. 1979, Ap\&SS, 65, 215
\bibitem[\protect\citeauthoryear{Hollenbach \& Salpeter}{1970}]{Holl70}Hollenbach, D., 
Salpeter, E. E., 1970, J. Chem. Phys., 53, 79
\bibitem[\protect\citeauthoryear{Hasegawa et al.}{1992}]{Hase92}Hasegawa, T., 
Herbst, E., Leung, C.M., 1992, APJ, 82, 167
\bibitem[\protect\citeauthoryear{Ioppolo et al.,}{2008}]{Ioppolo08} 
Ioppolo, S., Cuppen, H. M., Romanzin, C., Van Dishoeck, E. F., Linnartz, H.,
2008, APJ, 686, 1474
\bibitem[\protect\citeauthoryear{Jiang et al.,}{2000}]{Jiang00} 
Jiang, B. W., Szczerba, R., \& Deguchi, S., 2000, A\&A, 362, 273
\bibitem[\protect\citeauthoryear{Keane et al.,}{2001}]{Kean01}
Keane, J. V., Boogert, A. C. A., Tielens, A. G. G. M., Ehrenfreund, P., 
Schutte, W. A., 2001, A\&A, 375L, 43
{\it \bibitem[\protect\citeauthoryear{Klemm et al.,}{1975}]{Klem75}
Klemm, R. B., Payne, W. A., Stief, L. J. 1975, in Chemical Kinetic Data for
the Upper and Lower Atmosphere, ed. S. W. Benson ( New York: Wiley),61}
\bibitem[\protect\citeauthoryear{Lee et al.,}{1996}]{Lee96} 
Lee, H., H., Herbst, E., Pineau des Forets, G., Roueff, E., Le Bourlot, J., 1996, A\&A, 311, 690 
\bibitem[\protect\citeauthoryear{Leger et al.,}{1985}]{Leger85} 
Leger, A., Jura, M., Omont, A., 1985, A\&A, 144, 147
\bibitem[\protect\citeauthoryear{Leitch \& Williams}{1985}]{Leitch85} 
Leith-Devlin, M., A.,  Williams, D., A., 213, 295, MNRAS, 1985
\bibitem[\protect\citeauthoryear{Melius \& Blint}{1979}]{Mel79} 
Melius, C. F., \& Blint, R. J. 1979, Chem. Phys. Lett., 64, 183
\bibitem[\protect\citeauthoryear{Murakawa et al.,}{2000}]{Mur00} 
Murakawa, K., Tamura, M., Nagata, T., 2000, APJSS, 128, 603 
\bibitem[\protect\citeauthoryear{Nguyen et al.,}{2002}]{Ngu02} 
Nguyen, T. K., Ruffle, D. P., Herbst, E., Williams, D. A. 2002, MNRAS, 329,301 
\bibitem[\protect\citeauthoryear{Onishi et al.,}{1996}]{Onish96} 
Onishi, T., Mizuno, A., Kawamura, A., Ogawa, H., \& Fukui, Y. 1996, ApJ, 465, 815
\bibitem[\protect\citeauthoryear{Roberts \& Herbst}{2002}]{Robert02} 
Roberts, H., \& Herbst, E., 2002, A\&A, 395, 233
\bibitem[\protect\citeauthoryear{Schiff}{1973}]{Schiff73} 
Schiff, H., I., 1973, in physics and chemistry of Upper Atmospheres, ed. B. M.
McCormac ( Dordrecht: Reidel ), 85 
\bibitem[\protect\citeauthoryear{Shalabiea et al.,}{1994}]{Shala94} 
Shalabiea, O. M., greenberg, J. M., 1994, A\&A, 290, 266
\bibitem[\protect\citeauthoryear{Stantcheva et al.}{2002}]{Stan02}Stantcheva, T., 
Shematovich, V. I., Herbst, E., 2002, A\&A, 391, 1069
\bibitem[\protect\citeauthoryear{Talbi et al.}{2002}]{Talbi06}
Talbi, D., Chandler, G. S., Rohl, A. L., 2006, Chem. Phys., 320, 214
\bibitem[\protect\citeauthoryear{Tielens \& Hagen}{1982}]{Tiel82} Tielens, A. G. G. M., 
Hagen, W., 1982, A\&A, 114, 245
\bibitem[\protect\citeauthoryear{Vandenbussche et al.,}{1999}]{Vand99} 
Vandenbussche, B., Ehrenfreund, P., Boogert, A., C., A., van Dishoeck, E., F., Schutte, W., A., 
Gerakines, P, A., Chiar, J., Tielens, A., G., G., M., Keane, J., Whittet, D., C., B., 
and 2 coauthors, 1999, A\&A, 346L, 57
\bibitem[\protect\citeauthoryear{Westley et al.,}{1995}]{West95} 
Westley, M. S., Baragiola, R. A., Johnston, R. E., Baratta, G., A., 1995, Nat, 373, 405
\bibitem[\protect\citeauthoryear{Watson \& Salpeter}{1972}]{Wat72}Watson, W. D., 
Salpeter, E. E., 1972, ApJ, 174, 321
\bibitem[\protect\citeauthoryear{Whittet et al.,}{1996}]{Whit96} 
Whittet, D. C. B., et al., 1996, A\&A, 315, L357
\bibitem[\protect\citeauthoryear{Woodall et al.,}{2007}]{Wood07} 
Woodall, J.,Agundez, M., Markwick Kemper, A. J., Millar, T. J.,
2007, A\&A, 466, 1197
\end{thebibliography}
\end{document}